\documentclass[a4paper,11pt]{article}
\pdfoutput=1
\usepackage{jheppub}

\usepackage{amsmath,amssymb}
\usepackage{amsmath}
\usepackage{booktabs}
\usepackage{braket}
\usepackage{enumitem}
\usepackage{float}
\usepackage{graphicx}
\usepackage{hhline}
\usepackage[utf8x]{inputenc}
\usepackage{slashed}
\usepackage[dvipsnames]{xcolor}
\usepackage{xspace}

\newcommand{\refeq}[1]{eq.~(\ref{eq:#1})}
\newcommand{\refeqs}[2]{eqs.~(\ref{eq:#1})--(\ref{eq:#2})}
\newcommand{\reffig}[1]{figure~\ref{fig:#1}}
\newcommand{\refsec}[1]{section~\ref{sec:#1}}
\newcommand{\reftab}[1]{table~\ref{tab:#1}}

\newcommand{\GeV}{\text{GeV}}
\newcommand{\MeV}{\text{MeV}}

\renewcommand{\Im}{\operatorname{Im}}
\newcommand{\order}[1]{\mathcal{O}\left(#1\right)}

\newcommand{\para}{\parallel}

\allowdisplaybreaks

\title{Revisiting $B \to \pi\pi \ell \nu$ at Large Dipion Masses}

\author[1]{Thorsten Feldmann,}
\emailAdd{thorsten.feldmann@uni-siegen.de}

\author[2]{Danny van Dyk,}
\emailAdd{danny.van.dyk@gmail.com}
\author[1]{Keri Vos}
\emailAdd{keri.vos@uni-siegen.de}
\affiliation[1]{%
    Theoretische Physik 1, Universit\"at Siegen, 
    Walter-Flex-Stra\ss{}e 3, D-57068 Siegen, Germany%
}
\affiliation[2]{%
    Physik Department, Technische Universit\"at M\"unchen, James-Franck-Stra\ss{}e 1, D-85748 Garching, Germany
}

\abstract{We revisit QCD factorization of $B\to \pi\pi$ form factors at large
dipion masses, by deriving new constraints based on the analyticity properties
of these objects.
We then propose a parametrization of the form factors, inspired by the
leading-twist QCD factorization formula, that incorporates all known analytic properties.
This parameterization is used to interpolate between the QCDF results and the constraints from the $B^*$ pole. Based on this interpolation, we predict the $B\to \pi\pi\ell\nu$ decay rate
in a larger phase space region than previous studies could.
We obtain a partially-integrated branching ratio up to $\mathcal{B} \simeq \order{10^{-6}}$, which implies that
a measurement of the non-resonant semileptonic decay is potentially within reach of the Belle II experiment.
}


\note{Preprint: EOS-2018-01, SI-HEP-2018-23, QFET-2018-14, TUM-HEP-1149/18}

\begin{document}

\maketitle

\section{Introduction}
\label{sec:intro}

Semileptonic $b\to u$ transitions are used to determine the Standard Model (SM) parameter
$|V_{ub}|$. Contemporary determinations based on data provided by the B-factory experiments
BaBar and Belle, as well as the LHCb experiment show tensions between determinations from
exclusive and inclusive decays. Among the former, the decay $B\to \pi\ell\nu$ provides presently
the most control over the hadronic matrix elements, whose knowledge is required for the
$|V_{ub}|$ determinations. The matrix elements are known from lattice QCD
\cite{Flynn:2015mha,Lattice:2015tia,Colquhoun:2015mfa} and Light-Cone Sum Rules
(LCSRs) \cite{Duplancic:2008ix,Imsong:2014oqa,Wang:2015vgv}. In addition, the decay $B\to \rho\ell\nu$ is widely discussed as a further promising
channel. However, the $\rho$ is not an asymptotic state, since it decays rapidly through the
strong interaction. Alternative exclusive determinations of $|V_{ub}|$ or constraints on effects due
to physics beyond the SM (BSM) can be obtained from e.g., the three-body decay $B_s\to K\ell\nu$ \cite{Khodjamirian:2017fxg,Lattice:2017vqf};
the effective three-body decay $B_s\to K^*(\to K\pi)\ell\nu$ \cite{Feldmann:2015xsa} (with a much narrower
state $K^*$ compared to the rather wide $\rho$); and the four-body decay $B\to \pi\pi\ell\nu$
\cite{Faller:2013dwa,Kang:2013jaa,Meissner:2013hya,Boer:2016iez,Cheng:2017sfk,Cheng:2017smj,Hambrock:2015aor}.
In this article we will focus on a study of the hadronic matrix elements for the four-body decay,
which are also an important input to studies of the fully-hadronic decay $B\to \pi\pi\pi$
\cite{Krankl:2015fha,Klein:2017xti}.\\

A QCD Factorization (QCDF) formula for $B\to \pi\pi$ form factors at large dipion masses
was proposed in reference \cite{Boer:2016iez}. A major drawback to its phenomenological
applications are the phase-space limits that are needed to ensure the factorization into
the soft $B\to \pi$ form factor and pion and $B$-meson Light-Cone Distribution Amplitudes (LCDAs).
Indeed, in reference \cite{Boer:2016iez} the partially-integrated branching ratio of $B\to \pi\pi\ell\nu$
decays was found to range from $4\cdot 10^{-13}$ to $4\cdot 10^{-10}$, depending on the applied phase-space cuts. Given the expected size of the Belle II data set, the smallness of the branching ratio draws into question the prospects for measurements in the QCDF-accessible phase space.\\

Due to the large number of independent kinematic variables in the four-body decay $B\to \pi\pi\ell\nu$,
the analytic structure of the $B\to \pi\pi$ form factors is somewhat more complicated than for
the $B\to \pi$ or $B\to \rho$ form factors. Hadronic intermediate states, which may contribute 
to the $B \to \pi\pi \ell\nu$ decay in a dual way compared to the LO partonic picture, include:
\begin{description}
  \item[$\boldsymbol{B \to R_n(\to \pi\pi) \ell \nu}$] The $R_n = \rho,f_0,\rho',\dots$ are the light 
    resonances arising from a branch cut in the variable $k^2$. For a
    point-wise description of the $k^2$ spectrum detailed knowledge of the
    cut is required. However, in this work we focus on more inclusive observables
    far above the $\pi\pi$ threshold, which do not resolve the individual $R_n$ resonances.
  
  \item[$\boldsymbol{B \to X_b (\to \ell \nu) \pi \pi}$] Here, the $X_b = B^*, B_1, B$ is either the vector,
  axialvector or pseudoscalar resonance, respectively, that contributes to the $q^2$ spectrum
  of the individual form factors.
  
  \item[$\boldsymbol{B \to B^* (\to \pi_2 \ell \nu) \pi_1}$] The heavy vector meson $B^*$ contributes
    here as a resonance in the $\hat{q}^2 \equiv (p-k_2)^2=(q+k_1)^2$ spectrum.
    The aim of this article is to improve our description of the $B\to \pi\pi$ form factors
    by including \emph{this} resonant contribution and to thereby extend the form factors' reach.
\end{description}
Our parametrization makes use of two conformal maps to accelerate convergence. The use of
residues to improve our understanding of the form factors is an approach known from prior
phenomenological studies, such as those deriving unitarity bounds for $b\to c$ form factors \cite{Boyd:1994tt,Caprini:1997mu} and $b\to u$ 
form factors \cite{Bharucha:2010im}.\\

To this extent, we introduce the following independent kinematic variables,
\begin{equation}
  k^2 = (k_1+k_2)^2 \,, \qquad 
  q^2 = (q_1+q_2)^2 \,, \qquad 
  \hat q^2\equiv(p-k_2)^2=(q+k_1)^2 \,,
\end{equation}
with $q^2 \leq (M_B - \sqrt{k^2})^2$ and $\hat q^2 \leq M_B^2 - k^2$.
In the phase space of interest -- i.e. the phase space for the semileptonic $B$ decay --
the amplitudes are thus analytic functions of $q^2$ and $\hat q^2$, while they
exhibit a branch cut in $k^2$ that gives rise to resonant $\rho$, $f_0$, and
similar contributions. It then makes sense to perform the usual $z$-expansion, trading 
\begin{align}
  q^2 & \qquad  \mbox{for} \quad  z \equiv z(q^2) =   
  \frac{\sqrt{t_+-q^2}-\sqrt{t_+-t_0}}{\sqrt{t_+-q^2}+\sqrt{t_+-t_0}} \,, 
  \\
  \hat q^2 & \qquad \mbox{for} \quad \hat z \equiv \hat z(\hat q^2) =  
  \frac{\sqrt{\hat t_+-\hat q^2}-\sqrt{\hat t_+-t_0}}{\sqrt{\hat t_+-\hat q^2}+\sqrt{\hat t_+- t_0}} \,,
\end{align}
where
\begin{align}
  t_+       & = (M_B + 2 M_\pi)^2\,, &    t_0       & = 0\,,\\
  \hat{t}_+ & = (M_B + M_\pi)^2\,, &    \hat{t}_0 & = \hat{t}_+ - \sqrt{\hat{t}_+}\, \sqrt{\hat{t}_+ - M_{B^*}^2}\,.
\end{align}
Here $t_+$ and $\hat t_+ $ denote the thresholds of the hadronic continuum
in the respective channels and $t_0$ and $\hat{t}_0$ are the reference values that we will use for the \emph{extrapolation} of the perturbative result to larger values of $q^2$ and $\hat q^2$.
The expression for a generic dipion form-factor function will thus be given as
\begin{align}
  F(k^2,q^2,\hat q^2) & \sim \sum_{i,j} z^i \, \hat z^j \, f_{ij}(k^2) \,.
\end{align}
up to kinematic prefactors that arise from the definition of the form factors.
The concrete parametrization that we propose, supplemented by appropriate ``Blaschke factors'' to
account for subthreshold $B$-meson resonances, is discussed in \refsec{param:qcdf}. We continue
with our results for the $B^*$-pole residues in \refsec{param:analyticity}. Our numerical results
then follow in \refsec{results}, before we conclude in \refsec{conclusion}.

\section{Parametrization of the $B\to \pi\pi$ Form Factors and Theoretical Constraints}
\label{sec:param}

In the following we provide the rationale for our proposed parametrization, and provide
constraints on its parameters based on two theoretical results. Common to the following
discussions are the use of a basis of Dirac structures that define the $B\to \pi\pi$ form
factors. Throughout we will use the basis of \cite{Faller:2013dwa}, which reads:
\begin{equation}
\label{eq:Gamma-basis}
\begin{aligned}
\Gamma_t     & = \frac{-\slashed{q} \gamma_5}{\sqrt{q^2}}                     &  \Gamma_0     & = \frac{2\sqrt{q^2} \slashed{k}_0 \gamma_5}{\sqrt{\lambda_B}} \\
\Gamma_\para & = \frac{\slashed{\bar k}_\para \gamma_5}{\sqrt{k^2} \sin^2 \theta_\pi} &  \Gamma_\perp & = \frac{i\slashed{\bar q}_\perp}{\sqrt{k^2} \sin^2 \theta_\pi}         \, \ ,
\end{aligned}
\end{equation}
where $\lambda_B = \lambda(q^2, k^2, M_B^2)$ is the K\"all\'en function.

\subsection{Inspiration by the QCD Factorization Formulas}
\label{sec:param:qcdf}

We set out to produce a parametrization of the dipion form factor that is compatible
with the QCDF formula at large dipion masses, but which can also be augmented with
further constraints. To obtain better insight into the analytic dependence on the
kinematic variables, we start from the QCDF expressions to leading order in $\alpha_s$
and to leading twist as given in \cite{Boer:2016iez}:
\begin{align}
 &\langle \pi^+(k_1) \pi^-(k_2) |\bar\psi_u \Gamma \psi_b|B^-(p)\rangle 
 \cr 
 = & \frac{2\pi f_\pi}{k^2} \, \xi_\pi(\hat q^2) \, 
 \int_0^1du \, \phi_\pi(u) \, T_\Gamma^{\rm I}(u,k^2,q^2,\hat q^2) + \ldots 
\end{align}
for any current $\Gamma$. The function $T_\Gamma^{\rm I}$ encodes the perturbative
description of the dynamics related to the hadronic subprocesses 
$Y_b \to \pi^+ \ell^- \bar \nu$ in 
\begin{equation}
  B^-(p) \longrightarrow \left[Y_b(\hat q) \to \pi^+(k_1) \ell^-(q_1) \bar \nu(q_2)\right]  \pi^-(k_2) \,.
\end{equation}
The simplest
contributing hadronic states are given by
\begin{equation}
  (i)\,: \quad Y_b = B^{*} \,, \qquad (ii)\,: \quad Y_b = B^{(*)} \pi \,. 
\end{equation}
They imply a simple pole at $\hat q^2 = M_{B^*}^2$ and a cut for $\hat q^2 \geq (M_B^{(*)}+M_\pi)^2$.
Notice that these should be taken care of by a realistic implementation of the $B^- \to \pi^-$
form factor as a function of $\hat q^2$, respectively by its $\hat z$-expansion.

We investigate which terms are required in our parametrization by taking
a closer look at the results for $\Gamma = \Gamma_\perp$ as an example:
\begin{align}
\label{eq:param:Tperp-QCDF}
\frac{1}{\sqrt{k^2}} \,   T_\perp^{\rm I} = \sqrt{\lambda_B} \, 
\frac{i \alpha_s C_F}{2N_C} \, \frac{f_2(u)}{M_B^2-\hat q^2} \,,
\end{align}
with 
\begin{align}
  f_1(u) &= \frac{k^2}{M_B^2 - \bar u \, q^2 - u \, \hat q^2} \,, \qquad 
  f_2(u) = \frac{1}{\bar u} \, \frac{M_B^2 - \hat q^2}{M_B^2 - \bar u \, q^2 - u \, \hat q^2} \,.
\end{align}

Concerning the $q^2$ and $\hat q^2$ dependence in the subsequent decay,
we may improve the description of the form factors by including
appropriate Blaschke factors for the $B^*$ resonance in both channels  
below the $B \pi\pi (B\pi)$ continuum thresholds,
\begin{align}
P_{B^{*}} &\equiv \frac{1-z \, z_{B^{*}}}{z-z_{B^{*}}} \,, \qquad z_{B^{*}} = z(M_{B^{*}}^2,t_0)  \, , \\
\hat P_{B^{*}} &\equiv \frac{1-\hat z \, \hat z_{B^{*}}}{\hat z-\hat z_{B^{*}}} \,, \qquad \hat z_{B^{*}} = \hat z(M_{B^{*}}^2,t_0)  \, ,
\end{align}
respectively. These are implemented as
\begin{align}
\label{eq:param:Tperp-expanded}
\frac{1}{\sqrt{k^2}} \,   T_\perp^{\rm I} = \sqrt{\lambda_B}
\, \frac{i \alpha_s C_F}{2N_C}
\, P_{B^*}   \hat P_{B^*} \, \sum_{i,j} z^i \, \hat z^j \, f^\perp_{ij}(k^2) \,.
\end{align}
Matching \refeq{param:Tperp-expanded} onto \refeq{param:Tperp-QCDF}, we obtain to first approximation
\begin{align}
\sum_{i,j} z^i \, \hat z^j \, f^\perp_{ij}(k^2) &\simeq 
\frac{1}{\bar u \, M_B^2} \left( 
 z \, (1-z_{B^*}^2) - z_{B^*} \, (1- 4\bar u \, z - 4 u \, \hat z)  
\right) + \ldots \ , 
\label{fperpij}
\end{align}
Notice that the $z$-expansion turns the integration over the quark momentum fraction $u$ 
in the convolution with the pion LCDA rather simple. It is also to be noticed that the 
$k^2$-dependence decouples from the convolution integral.\\

As mentioned in the introduction, various $X_b = B^*,B_1,B$ states contribute as one-body
hadronic intermediate states to the dispersive representation of the form factors as functions
of $q^2$. In principle, one would need to specify which exact resonant state contributes.
Since we work at very small values of $q^2 \ll \lbrace M_{B^*}^2, M_{B_1}^2, M_B^2\rbrace$
we replace the individual poles with one effective pole at $q^2 = M_{B^*}^2$ by means of the
Blaschke factor $P_{B^*}$. This choice corresponds to the resonant contribution in the $B\to \pi$
vector form factor $f_+$, which we use later on for numerical predictions of the QCDF formulas
and analyticity constraints.\\

The discussion above can be generalized to all four (axial)vector form factors. Our
parametrizations then read:
\begin{align}
    \label{eq:param:Ftime}
    F_{t}^{\text{param}}
        & \equiv P_{B^*} \, \hat P_{B^*}\, \frac{M_B^3}{k^2\,\sqrt{q^2}} \,  \bigg[a_t + b_t \frac{M_B^2 - k^2}{M_B^2} + c_t \frac{(M_B^2- k^2)^2}{M_B^4} \bigg]\,, \\
    \label{eq:param:Flong} 
    F_{0}^{\text{param}}
        & \equiv P_{B^*} \, \hat P_{B^*}\, \frac{M_B^5}{k^2\, \sqrt{q^2\, \lambda_B}} \,  \bigg[a_0 + b_0 \frac{M_B^2 - k^2}{M_B^2} + c_0 \frac{(M_B^2- k^2)^2}{M_B^4} \bigg]\,, \\
    \label{eq:param:Fperp}
    F_{\perp}^{\text{param}}
        & \equiv P_{B^*} \, \hat P_{B^*}\, \frac{\sqrt{\lambda_B}}{M_B\, \sqrt{k^2}} \,  \bigg[a_\perp + b_\perp \frac{M_B^2 - k^2}{M_B^2} + c_\perp \frac{(M_B^2- k^2)^2}{M_B^4} \bigg]\,, \\
    \label{eq:param:Fpara}
    F_{\para}^{\text{param}}
        & \equiv P_{B^*} \, \hat P_{B^*}\, \frac{M_B}{\sqrt{k^2}} \,  \bigg[a_\para + b_\para \frac{M_B^2 - k^2}{M_B^2} + c_\para \frac{(M_B^2- k^2)^2}{M_B^4} \bigg]\,.
\end{align}
By using appropriate overall normalizations that reflect the dominant kinematic dependence of the
QCDF results, we minimize the number of parameters needed later on in the fits.
For all polarizations $\lambda$ the coefficients $a_\lambda, b_\lambda, c_\lambda$ still
require expansion in both $z$ and $\hat{z}$. The precise type of expansion is not
relevant at this point, and will be discussed detail in \refsec{results}.
Constraints on the $B^*$ pole in the variable $\hat{z}$ can be readily included by replacing
$\hat{P}_{B^*}$ with its residue:
\begin{equation}
    \lim_{\hat q^2 \to M_{B^*}^2} (q^2 - M_{B^*}^2)\,\hat P_{B^*} = 4 (M_{B^*}^2 - \hat t_+)\,.
\end{equation}

The QCDF expressions give good control over the behaviour in $\hat z$ for $0.25 \lesssim \hat z \lesssim 0.40$.
However, our aim is to extrapolate from the region where QCDF is applicable to the
larger $B\to \pi\pi$ phase space. This is achieved by imposing additional constraints at $\hat{z}<0$, which
are obtained from the $B^*$ pole, as discussed in the next section. Our approach is best illustrated
using $\cos\theta_\pi$ rather than $\hat{z}$. At small values of $k^2 \sim 7\,\GeV^2$ the QCDF predictions are
limited to the phase space $|\cos\theta_\pi| < 1/3$. The $B^*$-pole in the variable $\hat{q}^2$ then ``lives''
at unphysical values of $\cos\theta_\pi \sim 2$. These two constraints then anchor our parameterization on
both sides of the QCDF-inaccessible phase space $1/3 < \cos\theta_\pi \leq 1$, therefore turning an
extrapolation problem into an interpolation.\\

\subsection{Analyticity Constraints}
\label{sec:param:analyticity}

The one-body contributions to the dispersion relation of the $B^0\to \pi^+\pi^-$ form factors
in the variable $\hat q^2$ yield:
\begin{align}
  F_\lambda(q^2, \hat q^2, k^2)
      & = \frac{1}{2\pi} \sum_{\lambda'}\int_{\hat t_+}^\infty d\hat t \, \frac{\braket{\pi^+ | \bar{q}\,\Gamma_\lambda b | B^*}\,\braket{B^{*+}\pi^- | B^0}}{\hat t - \hat q^2} 2\pi \delta(\hat t - M_{B^*}^2)\\
  \nonumber
      & = \frac{-1}{2\pi} \int_{\hat t_+}^\infty d\hat t \, \frac{2\pi \delta(\hat t - M_{B^*}^2) \, \xi_{B^*\to \pi}(q^2) \, g_{B^*B\pi}}{\hat t - \hat q^2}
      M^{\alpha}(\lambda) p^\beta \left[-g_{\alpha\beta} + \frac{\hat{q}_\alpha \hat{q}_\beta}{\hat q^2}\right]\,.
\end{align}
Here $\xi_{B^*\to \pi}$ refers to the soft form factor in $B^*\to \pi$ matrix elements,
defined in complete analogy to the $B\to \pi$ form factors in the SCET limit \cite{Beneke:2000wa}:
\begin{align}
  \langle \pi^+(k_1)| \bar{q} \,\Gamma_\lambda b |B^*(\hat q,\lambda')\rangle
      & =  \xi_{B^*\to \pi}(q^2 = (\hat q - k_1)^2) \, {\rm tr}\left[ \slashed{k}_1  \gamma_5 \, \Gamma_\lambda \, \frac{1+\slashed v}{2} (-i\gamma^\alpha) \right] \epsilon_\alpha(\lambda')\nonumber \\
        & \equiv \xi_{B^*\to \pi}(q^2) M^\alpha(\lambda) \, \epsilon_\alpha(\lambda')\,,
\end{align}
where $\lambda'$ is the $B^*$ polarization. In addition, we use the $B^*B\pi$ coupling \cite{Khodjamirian:1999hb}
\begin{align}
  \braket{B^{*+}(\hat q,\lambda') \pi(k_2)|B^{0}(p)}
      & = -(p \cdot \epsilon^*(\hat q,\lambda')) \, g_{B^*B\pi}\,,
\end{align}
as well as the completeness relations
\begin{align}
	\sum_{\lambda'} \epsilon_\mu(\hat{q}; \lambda')\, \epsilon_\nu^*(\hat{q}; \lambda')\, = -g_{\mu\nu} + \frac{\hat{q}_\mu \hat{q}_\nu}{\hat{q}^2}\,.
\end{align}

Our aim is now to relate the residue of the $B^*$ pole to the residue of our parametrization,
therefore anchoring it at large values of $\hat{q}^2 = M_{B^*}^2$. We determine the
imaginary part of the residue on the $B^*$ pole to be
\begin{align}
  \Im \left[\operatorname{Res}_{\hat{q}^2 \to M_{B^*}^2} F_\lambda(q^2, \hat{q}^2, k^2)\right]
      & = \Im \left[\lim_{\hat{q}^2 \to M_{B^*}^2} (\hat{q}^2 - M_{B^*}^2) F_\lambda(q^2, \hat{q}^2, k^2) \right] \nonumber \\
      & = \xi_{B^*\to\pi}(q^2) \; g_{B^*B\pi} \; \Im \Big[S_\lambda(q^2, \hat q^2 = M_{B^*}^2, k^2)\Big]\,.
\end{align}
where
\begin{equation}
    S_\lambda \equiv M^\alpha(\lambda) p^\beta \left[-g_{\alpha\beta} + \frac{\hat{q}_\alpha \hat{q}_\beta}{M_{B^*}^2}\right]\,.
\end{equation}
The soft form factor $\xi_{B^*\to\pi}$ is not well known. However, using heavy quark symmetry it
can be related to the soft form factor $\xi_{B\to \pi}$, which can be identified with the $B\to \pi$ vector form factor $f_+$ \cite{Beneke:2000wa}. Throughout we use the BCL parametrization for
the form factor \cite{Bourrely:2008za} with parameter values obtained from a LCSR study \cite{Imsong:2014oqa}.
The relation between the soft form factors is subject to power corrections,
which we estimate to be of the order of $30\%$.\\

For the different $\lambda$ polarization, we obtain 
\begin{align}
  S_t     & = -i\,\frac{M_B^2 (M_B^2 - M_{B^*}^2) (M_{B^*}^2 - q^2) - k^2 M_{B^*}^2 (M_B^2 + M_{B^*}^2)}{2 M_B M_{B^*}^2 \sqrt{q^2}}\,, \\
  S_0     & = -i\,\frac{\left(k^2 (M_B^2 + M_{B^*}^2) - (M_B^2 - q^2) (M_B^2 - M_{B^*}^2)\right) \left(k^2 M_{B^*}^2 + M_B^2(q^2 - M_{B^*}^2)\right)}{2 M_B M_{B^*}^2 \sqrt{q^2} \sqrt{\lambda_B}}\,. \\
  S_\perp & = -i\,\frac{\sqrt{k^2} \sqrt{\lambda_B} (M_B^2 + M_{B^*}^2)}{4 M_B M_{B^*}^2}\,, \\
  S_\para & = -i\,\frac{\sqrt{k^2}(M_B^4 + M_{B^*}^2(q^2 -k^2) + M_B^2(q^2 -3M_{B^*}^2-k^2))}{4M_B M_{B^*}^2}\,, 
\end{align}

Finally, we equate the two different expressions for the residues through the statistical procedure outlined in \refsec{results}.

\section{Phenomenological Applications}
\label{sec:results}

We proceed in three steps. First, we produce a theoretical likelihood that incorporates information
from the QCD factorization formulas as well as from the analyticity constraints.
Second, we discuss the concrete parametrization and provide results for the parameters
from a fit to the theoretical likelihood.
Third, we produce numerical estimates of two integrated $B\to \pi\pi\ell\nu$ observables in various phase-space bins.

\subsection{Theoretical Likelihood}
\label{sec:results:llh}

We use the QCDF expressions for the $B\to \pi\pi$ form factors to leading-order in $\alpha_s$
and to leading- and next-to-leading twist accuracy to produce synthetic data points. We generate
these data points for the form factors at the following values of the kinematic variables $k^2$, $q^2$, and $\cos\theta_\pi$:
\begin{align}
  k^2 & = 25\,\GeV^2 &
  q^2 & = 0.05\,\GeV^2 &
  \cos\theta_\pi & \in \lbrace -1.0, -0.5, 0, +0.5, +1.0\rbrace\\
  k^2 & = 19\,\GeV^2 &
  q^2 & \in \lbrace 0.05, 0.60 \rbrace\,\GeV^2 &
  \cos\theta_\pi & \in \lbrace -1.0, -0.5, 0, +0.5, +1.0\rbrace\\
  k^2 & = 16\,\GeV^2 &
  q^2 & \in \lbrace 0.05, 1.50 \rbrace\,\GeV^2 &
  \cos\theta_\pi & \in \lbrace -1.0, -0.5, 0, +0.5, +1.0\rbrace\\
  k^2 & =  7\,\GeV^2 &
  q^2 & \in \lbrace 0.05, 1.50, 6.00 \rbrace\, \GeV^2 &
  \cos\theta_\pi & \in \lbrace -0.33, 0, +0.33 \rbrace\,,
\end{align}
for a total of $34$ QCDF data points per form factor.
The smallest value of $q^2$ was chosen as roughly $\mathcal{O}(m_\mu^2)$, in order to
regularize a divergence of kinematic origin in the form factors $F_0$ and $F_t$.
Following \cite{Boer:2016iez} we do not use the QCDF factorisation results when
the pion energies in the B-meson rest frame falls below a threshold of $\sim 1.2\,\GeV$,
which corresponds to a maximal value of $|\cos\theta_\pi| \simeq 0.33$ at $k^2 = 7\,\GeV^2$.\\

In addition to the QCDF expressions, we also produce synthetic data points for the imaginary
part of the residues of the form factors on the $B^*$ pole. The theoretical expressions for
the residue of the form factors are provided in \refsec{param:analyticity}. Fixing
$\hat{q}^2 = M_{B^*}^2$ still leaves two free kinematic variables. We choose the following
values of $k^2$ and $q^2$:
\begin{align}
    k^2 & \in \lbrace 10, 13 \rbrace\,\GeV^2 &
    q^2 & \in \lbrace 0.05, 1.50 \rbrace\,\GeV^2\,,\\
    k^2 & = 7\,\GeV^2 &
    q^2 & \in \lbrace 0.05, 1.50, 6.00 \rbrace\,\GeV^2\,.
\end{align}
By including the residues in the fit only for small values of $k^2$ we stabilize the fit
and supplement information in the space region where the QCDF formulas lack predictive
power.\\

\begin{table}[t]
\centering
\renewcommand{\arraystretch}{1.2}
\resizebox{\textwidth}{!}{
\begin{tabular}{|c| c| c| c| c|}
    \hline
    parameter                                   & value/interval          & unit     & prior                          & source/comments\\
    \hline
    \multicolumn{5}{|c|}{QCD input parameter}\\
    \hline
    $\alpha_s(m_Z)$                             &  0.1184  $\pm$ 0.0007   & ---      & gaussian $@$ $68\%$            &  \cite{Beringer:1900zz}\\
    $\mu$                                       &  $M_B/2$ $\pm$ $M_B/4$  & \GeV     & gaussian$^\dagger$ $@$ $68\%$  &  \\
    $\overline{m}_{u+d}(2\,\GeV)$               &  7.8 $\pm$ 0.9          & \MeV     & uniform $@$ $100\%$            & see \cite{Imsong:2014oqa}\\
    \hline
    \multicolumn{5}{|c|}{hadron masses}\\
    \hline
    $M_B$                                       &  5279.58                & \MeV     & ---                            &  \cite{Beringer:1900zz}
     \\
    $M_{B^*}$                                   &  5324.65                & \MeV     & ---                            &  \cite{Beringer:1900zz}
     \\
    $M_\pi$                                     & 139.57                  & \MeV     & ---                            &  \cite{Beringer:1900zz}
     \\
    \hline
    \multicolumn{5}{|c|}{parameters of the pion DAs}\\
    \hline
    $f_\pi$                                     & $130.4$                 & \MeV     & ---                            &  \cite{Beringer:1900zz}\\
    $a_{2}^{\pi}(1\,\GeV)$                      & $[0.09, 0.25]$          & ---      & uniform $@$ $100\%$             &  \cite{Khodjamirian:2011ub}\\
    $a_{4}^{\pi}(1\,\GeV)$                      & $[-0.04, 0.16]$         & ---      & uniform $@$ $100\%$     &  \cite{Khodjamirian:2011ub}\\
    $\mu_\pi(2\,\GeV)$                          & 2.5 $\pm$ 0.3           & \GeV     & ---                            &  $M_\pi^2/(\overline{m}_{u+d})$ \\
    \hline
    \multicolumn{5}{|c|}{hadronic couplings}\\
    \hline
    $g_{B^*B\pi}$                               & $30\pm 5$               & ---      & gaussian $@$ $68\%$     &  \cite{Imsong:2014oqa}\\
    \hline
\end{tabular}
}
\caption{The input parameters used in our numerical analysis.
We express the prior distribution as a product of
individual priors
that are either uniform or gaussian. The uniform priors cover the stated
intervals with
100\% probability. The gaussian priors cover the stated intervals with 68\%
probability,
and the central value corresponds to the mode of the prior. For practical
purposes,
random variates of the gaussian priors are only drawn from their respective 99\%
probability intervals. The prior for the parameters describing the $B\to \pi$
form factor $f_+$ are not listed here and taken from \cite{Imsong:2014oqa}.
$\dagger$: We restrict variates of the renormalization scale
$\mu$ to the interval $[M_B/4, M_B]$ to avoid unphysical values.
}
\label{tab:inputs}
\end{table}

For the production of all theory pseudo observables, both form factors and residues,
we closely follow \cite{Boer:2016iez}. We use the publicly available EOS \cite{EOS}
software, which already features a numerical implementation of the QCDF expressions
for the form factors. We extend EOS with an implementation of the $B^*$-pole residues.
To produce the pseudo data points we follow a Bayesian approach. Our choice of the a-priori
Probability Density Function (PDF) is summarized in \reftab{inputs}. We draw $10^6$
samples from the prior PDF, which are then used to produce the same number of samples for
each of the pseudo observables. The mean and parametric uncertainty are then estimated
through the sample mean and sample covariance of the pseudo observables.
Since both sets of predictions share a strong dependence on the value of the soft form
factor $\xi_\pi$, we find that all pseudo observables are strongly correlated, with some correlation
coefficients as large as $0.99$. However, we find that the covariance matrix is still regular.\\

Both the QCDF expressions for the form factors from \cite{Boer:2016iez} as well as our results for
the $B^*$-pole residues in \refsec{param:analyticity} only hold to leading power in an expansion
in $1/m_b$ and $1/E_\pi$, the pion energy in the $B$-meson rest frame. In order to account for
contributions beyond these leading-power expressions, we assign an ad-hoc systematic uncertainty of
$30\%$ of the central value. This systematic uncertainty is combined with the parametric uncertainty by
adding their respective covariance matrices.\\

Due to the large dimensionality, we provide the combined multivariate Gaussian likelihood
for all pseudo observables only in machine readable form, as part of the EOS software.
The total number of observations in the likelihood is $N_\text{obs} = 4 \times 41 = 164$.
The names of the newly added EOS constraints with $30\%$ added systematic uncertainty are
\begin{align*}
    &\texttt{B->pipi::F\_time[sys=0.30]@FvDV2018}  & \text{for} & \qquad F_t \\
    &\texttt{B->pipi::F\_long[sys=0.30]@FvDV2018}  & \text{for} & \qquad F_0\\
    &\texttt{B->pipi::F\_perp[sys=0.30]@FvDV2018}  & \text{for} & \qquad F_\perp\\
    &\texttt{B->pipi::F\_para[sys=0.30]@FvDV2018}  & \text{for} & \qquad F_\para
\end{align*}
respectively.

\subsection{Fit of the $B\to \pi\pi$ Form Factor Parameters}
\label{sec:results:params}

We now proceed to fit our expressions in \refeqs{param:Ftime}{param:Fperp} to the
theory likelihood constructed in \refsec{results:llh}. For this we need to define
explicitly the
expansion of the coefficient $a_\lambda$, $b_\lambda$ and $c_\lambda$ in $z$ and $\hat z$.
We use
\begin{align}
    x_\lambda
        & = \sum_{i=0}^3 x^\lambda_{0i} \, \hat{z}^i
        + \sum_{i=0}^2 x^\lambda_{1i} \, z\, \hat{z}^i\,,
\end{align}
for all polarisations $\lambda$ and all coefficients $x = a,b,c$.
This amounts to $21$ parameters per form factor, with $\theta$ representing the entire set of parameters. Our rationale for choosing these expansions is our power counting $\hat{z}^2 \simeq z$, and the fact
that we can achieve a good fit with the smallest number of parameters per form factor, as outlined below.\\

We carry out a fit to all form factor parameters $\theta$ simultaneously, which amounts
to a $84$ dimensional fit. We use Minuit2 to find the best-fit point of the posterior
PDF 
\begin{equation}
    P(\theta \, | \, \text{theory})
        = \frac{P(\text{theory} \, | \, \theta)\, P_0(\theta)}{Z}
\end{equation}
where $P_0(\theta)$ is the prior PDF, $P(\text{theory}\,|\,\theta)$ is the
likelihood, and $Z$ is the evidence. We find a total minimal $\chi^2 = 8.79$ for a total
of $N_\text{d.o.f.} = N_\text{obs} - N_\text{par.} = 164 - 84 = 80$ degrees of freedom.
The fit quality is therefore excellent, with a p value in excess of $99\%$.\\

\begin{figure}[H]
\centering
\begin{tabular}{cc}
    \includegraphics[width=.45\textwidth]{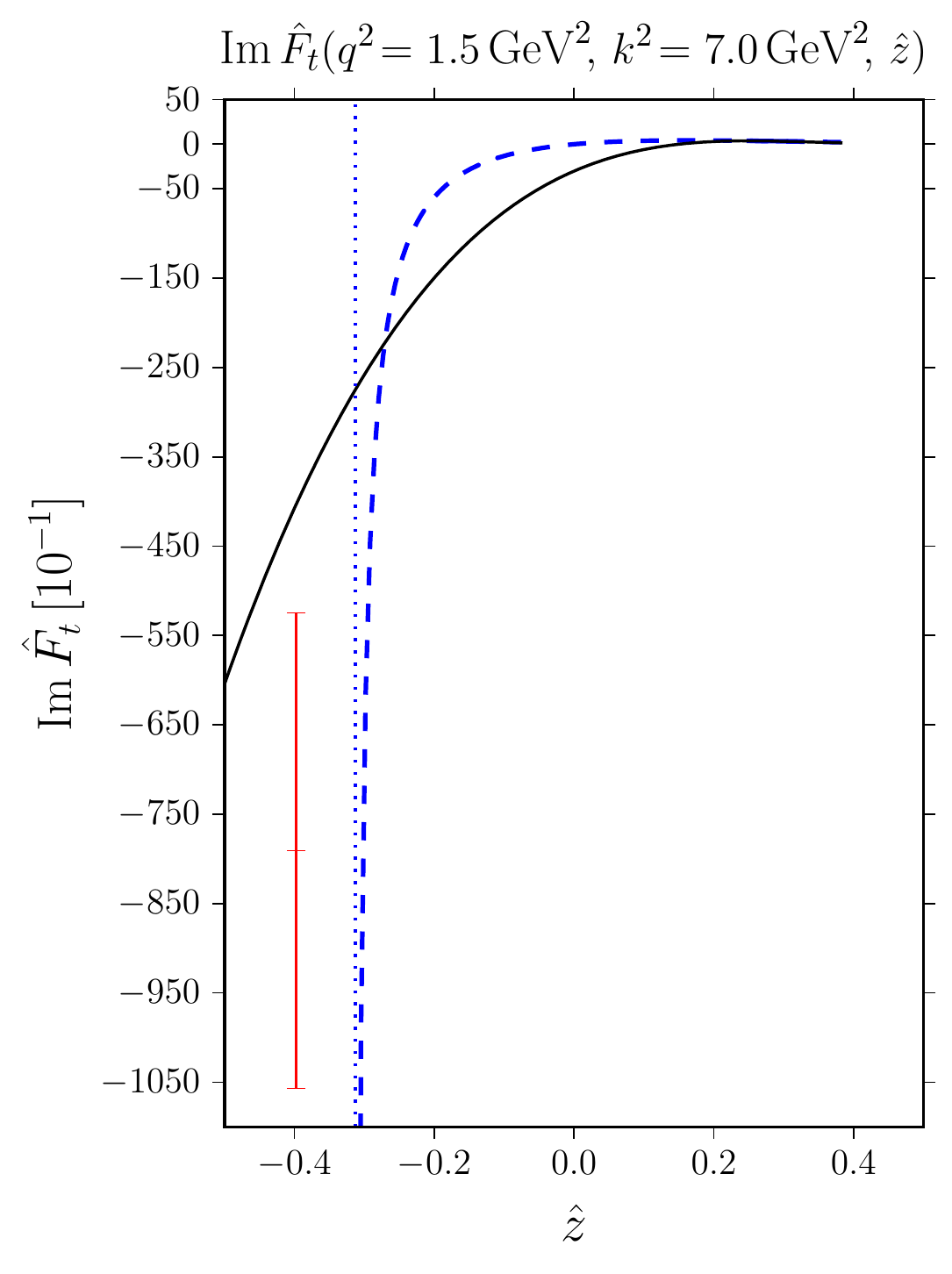} &
    \includegraphics[width=.45\textwidth]{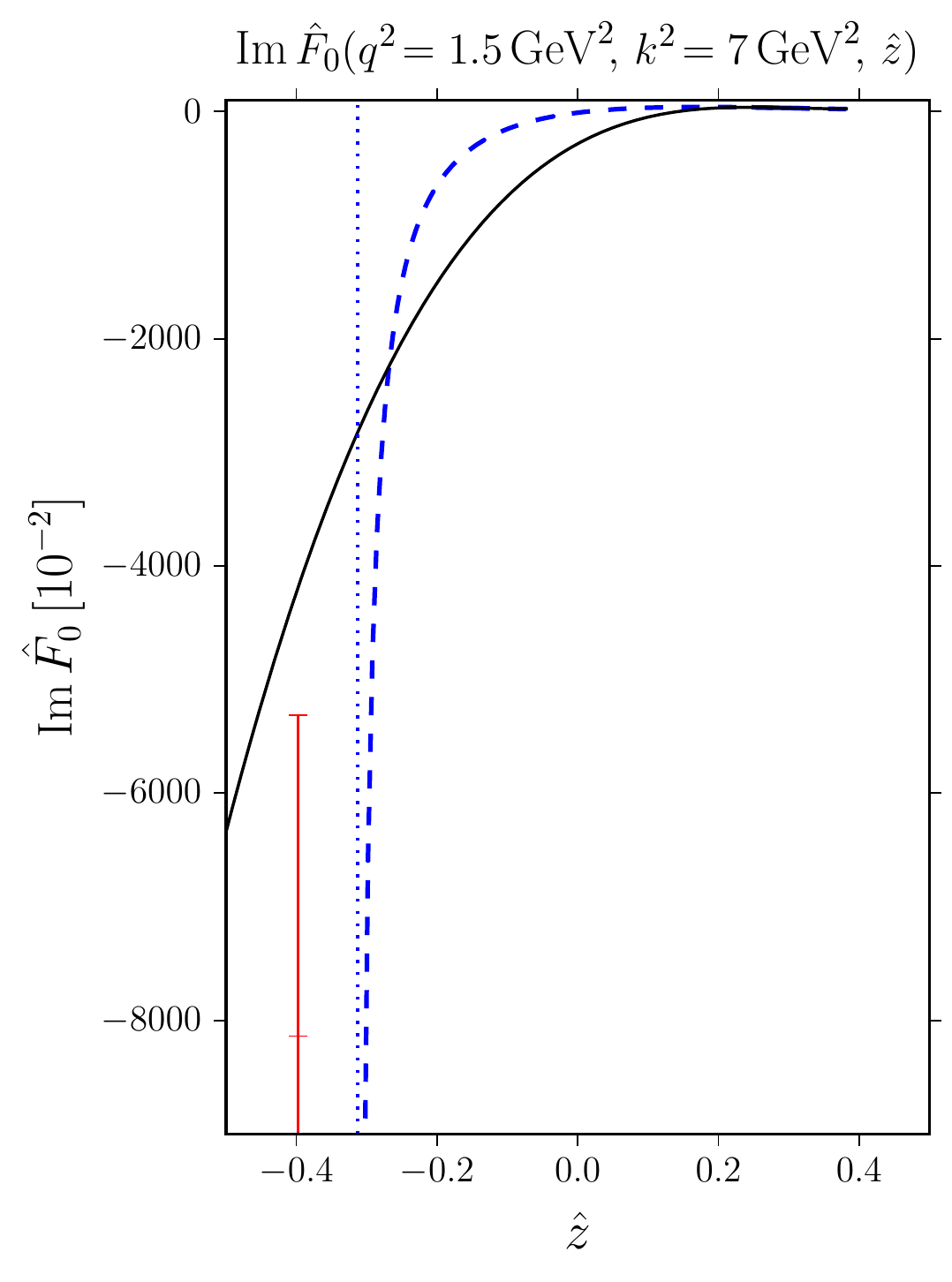}
    \\[-2em]
    \includegraphics[width=.45\textwidth]{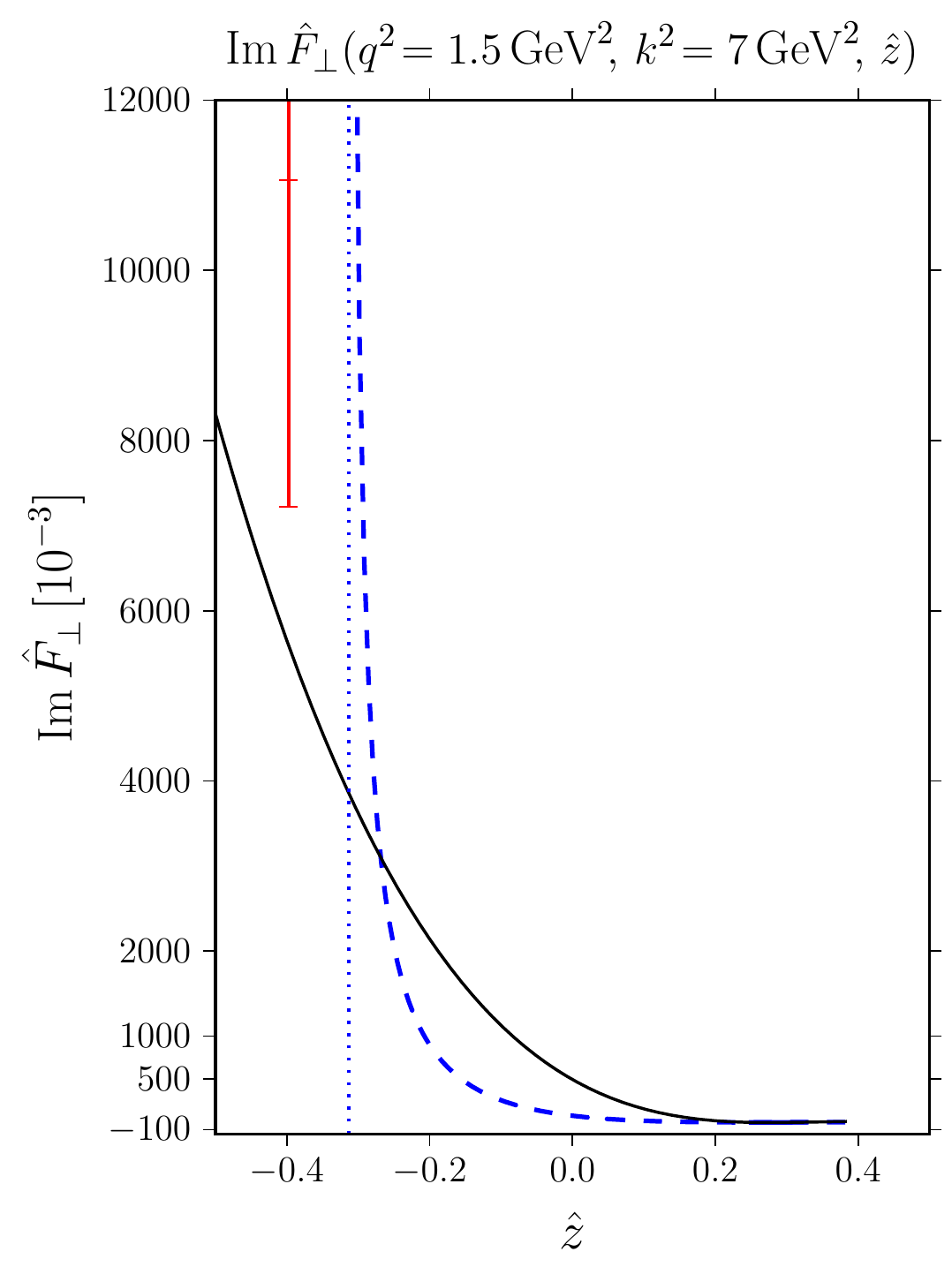} &
    \includegraphics[width=.45\textwidth]{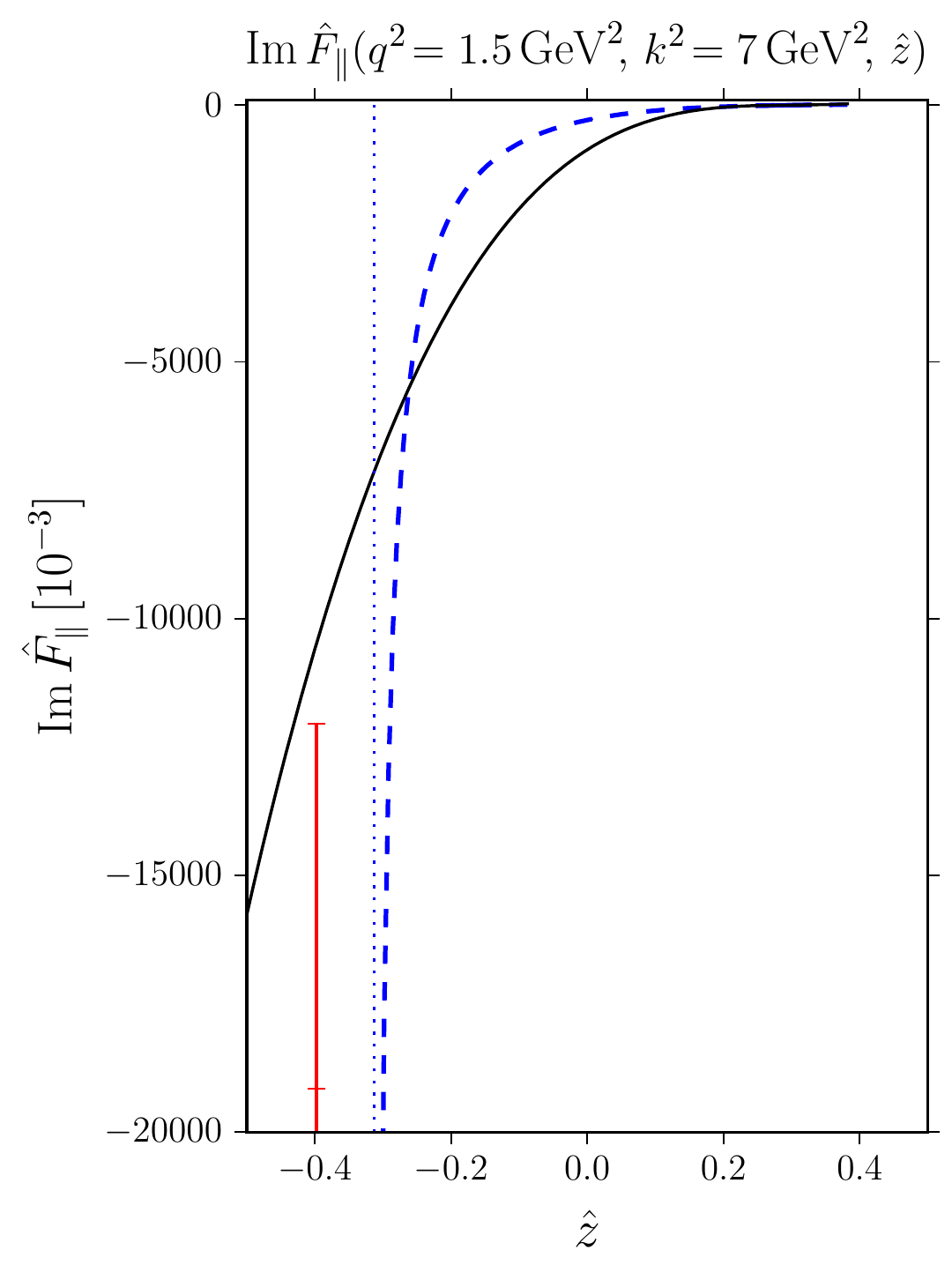}
\end{tabular}
\caption{
    Plots of the $\hat{z}$ dependence of the form factors
    in the phase space point $(q^2 = 1.5\,\text{GeV}^2, k^2 = 7\,\text{GeV}^2)$.
    Here $\hat{F}_\lambda \equiv F_\lambda / \hat{P}_{B^*}$ in order to be able
    to visualize the agreement between the parametrization and the residues on
    the $B^*$ pole.
    The dashed lines show the QCDF predictions at LO in $\alpha_s$, including both
    the twist-2 and twist-3 contributions. These predicitions are valid in the region where $0.25 \leq \hat{z} \leq 0.33$. Beyond this region, QCDF breaks down and the curve
    shown should be understood as merely an extrapolation.
    Our fit result, based on the parametrization in \refeqs{param:Ftime}{param:Fperp}, is shown as the black
    solid line. The residues following from \refsec{param:analyticity} are shown
    as red data points. The dotted vertical lines highlight $\hat{z}(\hat{q^2} = M_B^2)$,
    to show the unphysical pole emerging from the QCDF results
    at $\hat{q}^2 \simeq M_B^2$.
    With these plots we deliberately show the phase-space points that correspond to the largest
    tensions ($<2\sigma$) between the $B^*$ residues and our fit.
}
\label{fig:plots}
\end{figure}

We then use Minuit2's information on the parameters' uncertainties to inspire starting
values for the prior intervals. Our final prior intervals are then chosen to include at
least $99\%$ of their respective one-dimensional marginalized posteriors. For latter
purpose we produce $10^6$ posterior samples.
The sample mean and sample covariance of the posterior samples are provided as an EOS \cite{EOS}
constraint file in YAML format. The file is attached to the arXiv preprint of this
article as an ancillary file.
We show plots of the form factors $F_\lambda$, normalized to the Blaschke factor
$\hat{P}_{B^*}$, as functions of $\hat{z}$ and for fixed $q^2 = 1.5\,\GeV^2$ and
$k^2 = 7\,\GeV^2$ in \reffig{plots}. The singularity due to the unphysical $b$-quark
resonance in the QCDF results is clearly visible in the extrapolation of the
QCDF formulas (dashed lines).

\subsection{Numerical Results for $B\to \pi\pi\ell\nu$ Observables}
\label{sec:results:observables}

For the numerical illustration, we consider the three scenarios A, B and C discussed in  \cite{Boer:2016iez}.  These phase-space regions are defined as
\begin{gather}
    \tag{Region A}
    \begin{aligned}
        2 M_B^2 / 3 \simeq 18.6\,\GeV^2
            & \leq k^2 \leq M_B^2 \ ,\\
        0   & \leq |\cos\theta_{\pi}| \leq 1 \ .
    \end{aligned}\\[1em]
    \tag{Region B}
    \begin{aligned}
        M_B^2 / 2 \simeq 13.9\,\GeV^2
            & \leq k^2 \leq 2 M_B^2 / 3 \simeq 18.6\,\GeV^2 \ ,\\
        0   & \leq |\cos\theta_{\pi}| \leq 1 \ .
    \end{aligned}\\[1em]
    \tag{Region C}
    \begin{aligned}
        M_B^2 / 4 \simeq 7\,\GeV^2
            & \leq k^2 \leq M_B^2 \ ,\\
        0   & \leq |\cos\theta_{\pi}| \leq 1/3 \ .
    \end{aligned}
\end{gather}
In addition, we define three new scenarios which have an extended phase space compared to the previous ones.
\begin{gather}
    \tag{Region C'}
    \begin{aligned}
        M_B^2 / 4 \simeq 7\,\GeV^2
            & \leq k^2 \leq M_B^2 \ ,\\
        0   & \leq |\cos\theta_{\pi}| \leq 1 \ .
    \end{aligned}\\[1em]
    \tag{Region D}
    \begin{aligned}
        4\,\GeV^2
            & \leq k^2 \leq M_B^2 \ ,\\
        0   & \leq |\cos\theta_{\pi}| \leq 1/3 \ .
    \end{aligned}\\[1em]
    \tag{Region D'}
    \begin{aligned}
        4\,\GeV^2
            & \leq k^2 \leq M_B^2 \ ,\\
        0   & \leq |\cos\theta_{\pi}| \leq 1 \ .
    \end{aligned}
\end{gather}
For all regions $0.02\,\GeV^2 \leq q^2 \leq (M_B - \sqrt{k^2})^2$ holds. Region C' corresponds to region C
with an extended range for $|\cos\theta_\pi|$.
Regions D and D' are entirely new, and they correspond to an extrapolation in $k^2$
with respect to region C and C', respectively.
In \reftab{int-obs}, we present our results for two observables, the branching ratio $\mathcal{B}$ and the pionic forward-backward
asymmetry $A_\text{FB}^\pi$ as defined in \cite{Boer:2016iez}, in each of the specified regions. For comparison, we also include the results obtained in \cite{Boer:2016iez} for the regions A, B, and C.

\begin{table}[ht]
    \renewcommand{\arraystretch}{1.2}
    \centering
    \begin{tabular}{|c|c|c|c|c| r|}
        \hline
        & \multicolumn{2}{|c|}{result of \cite{Boer:2016iez}}
        & \multicolumn{2}{|c|}{our result}
        & \\
        phase space region & central & unc. & central & unc. & unit\\
        \hline
        \multicolumn{6}{|c|}{$\mathcal{B}(B^-\to \pi^+\pi^-\mu^-\bar\nu_\mu)\,/\,|V_{ub}|^{2}$}\\
        \hline
        (A)   & $ 2.93$ & $^{+1.00}_{-0.53}$ & $ 3.03$ & $^{ +1.20}_{-0.96}$ & $10^{-8}$\\
        (B)   & $ 9.31$ & $^{+3.23}_{-1.47}$ & $12.30$ & $^{ +3.90}_{-3.60}$ & $10^{-7}$\\
        (A+B) & $ 9.60$ & $^{+3.38}_{-1.52}$ & $13.30$ & $^{ +3.60}_{-3.90}$ & $10^{-7}$\\
        (C)   & $ 3.18$ & $^{+0.79}_{-0.71}$ & $ 4.88$ & $^{ +3.90}_{-1.70}$ & $10^{-5}$\\ \hline
        (C')  & ---     & ---                & $ 1.61$ & $^{ +1.80}_{-0.60}$ & $10^{-4}$\\
        (D)   & ---     & ---                & $ 0.74$ & $^{ +3.40}_{-0.48}$ & $10^{-2}$\\ 
        (D')  & ---     & ---                & $ 2.86$ & $^{+16.00}_{-1.80}$ & $10^{-2}$\\
        \hline
        \multicolumn{6}{|c|}{$A_\text{FB}^\pi(B^-\to \pi^+\pi^-\mu^-\bar\nu_\mu)$}\\
        \hline
        (A)   & $-1.96$ & $^{+0.16}_{-0.20}$ & $-1.72$ & $^{+1.70}_{-1.40}$ & $10^{-1}$\\
        (B)   & $-0.29$ & $^{+0.22}_{-0.22}$ & $+0.96$ & $^{+1.70}_{-1.30}$ & $10^{-1}$\\
        (A+B) & $-0.32$ & $^{+0.20}_{-0.24}$ & $+1.07$ & $^{+1.80}_{-1.40}$ & $10^{-1}$\\
        (C)   & $+1.25$ & $^{+0.08}_{-0.11}$ & $+6.31$ & $^{+0.99}_{-2.40}$ & $10^{-1}$\\ \hline
        (C')  & ---     & ---                & $+6.31$ & $^{+0.99}_{-2.40}$ & $10^{-1}$\\
        (D)   & ---     & ---                & $+6.82$ & $^{+0.90}_{-0.56}$ & $10^{-1}$\\ 
        (D')  & ---     & ---                & $+6.82$ & $^{+0.90}_{-0.56}$ & $10^{-1}$\\
        \hline
    \end{tabular}
    \renewcommand{\arraystretch}{1.0}
    \caption{
        Results for the partially-integrated branching ratio $\mathcal{B}$ (in units of $|V_{ub}|^2$)
        and the pionic forward-backward asymmetry $A_\text{FB}^\pi$ in different phase-space bins.
        \label{tab:int-obs}
    }
\end{table}

For the new regions C', D, and D', we obtain results that are significantly larger than those in
the QCDF regions A, B, and C. This is not surprising, since compared to region C, the new regions
span a phase space which is larger by factors of 3, $\sim$ 1.7, and $\sim$ 5, respectively.

Compared to region C', the partially-integrated branching ratios in regions D and D' exhibit large uncertainties. This is caused by our extrapolatation from data points at $k^2 \leq 7\,\GeV^2$ down
to $4\,\GeV^2$ for which we cannot expect that our QCDF-inspired parametrization
can still describe the form factors accurately. This can also be understood as a model error, since our
parametrization does not (and presently cannot) account for the light and broad dipion resonances
that contribute in the region $k^2 < 7\,\GeV^2$. Including these resonances might further improve our description of the form factors, which requires extensive further studies. We leave these for future work.\\

Exclusive determinations yield $|V_{ub}| \simeq 3.5\cdot 10^{-3}$. Assuming this value,
we find the $68\%$ probability intervals $[4, 5.1]\cdot 10^{-7}$ and $ [1.3, 23]\cdot 10^{-7}$ for the partially-integrated branching ratios in regions D and D', respectively.
We emphasize that this prediction indicates that a measurement at the Belle II experiment is feasible.
This shows that our strategy of including analyticity constraints is clearly beneficial, since it allows to consider a larger subset of the phase space.

\section{Conclusion}
\label{sec:conclusion}
We present an updated study of the form factors relevant for $B \to \pi\pi$ semileptonic decays, which were previously studied at large dipion masses in the framework of QCDF. These form factors feature interesting analytic properties.
Combining the QCDF results with information on the $B^*$ pole allows to interpolate the form factors
in the kinematic variable $\hat{q}^2$.
To this extent, we propose a parametrization that respects the dominant kinematic behavior of the QCDF formulas and
has a pole at the $B^*$ mass. Fitting this parametrization to all available predictions, we significantly extend the range of applicability of our theoretical framework.
The relevance of this is illustrated by larger values of the partially-integrated observables in the newly-defined
phase space region.
Our results indicate that, for moderately large
dipion masses, a phenomenological study of the $B\to \pi\pi \ell\nu$ decay with the upcoming Belle II data set is
feasible.

\acknowledgments

We are grateful to Frederik Beaujean for his help in finding and fixing a bug in
the statistics part of the EOS software.
We thank Philipp B\"oer for useful discussions in the early phase of this work.
T.F.~and K.K.V.~are supported by the Deutsche Forschungsgemeinschaft (DFG) within research unit FOR 1873 (QFET). D.v.D~is supported by the DFG within the Emmy Noether Programme under grant DY-130/1-1
and the DFG Collaborative Research Center 110 ``Symmetries and the Emergence of Structure in QCD''.

\bibliographystyle{JHEP-2}
\bibliography{references}

\end{document}